\documentclass[12pt]{article}

\usepackage[english]{babel}

\usepackage[a4paper,top=2cm,bottom=2cm,left=3cm,right=3cm,marginparwidth=1.75cm]{geometry}

\usepackage{amsmath}
\usepackage{graphicx}
\usepackage[colorlinks=true, allcolors=blue]{hyperref}
\usepackage{xcolor}
\usepackage[UTF8]{ctex}
\usepackage{multirow}
\usepackage{float}
\usepackage{makecell}
\title{Chiral Integrable Boundary States in the SU(4) Alternating Spin Chain}

\author{\small
	Yang Liu\textsuperscript{a}\thanks{\href{mailto:liuyang_2306@tju.edu.cn }{\texttt{liuyang\_2306@tju.edu.cn}}},
	\and \small
    Jun-Bao Wu\textsuperscript{a, b}\thanks{\href{mailto:junbao.wu@tju.edu.cn}{\texttt{junbao.wu@tju.edu.cn}}, corresponding author. }}

\date{\small
	\textsuperscript{a} Center for Joint Quantum Studies and Department of Physics, School of Science, Tianjin University, 135 Yaguan Road, Tianjin 300350, P. R. China \\
	\textsuperscript{b} Peng Huanwu Center for Fundamental Theory, 96 Jinzhai Road, Hefei, Anhui 230026, P. R. China 
	}

\begin{document}
\maketitle

\begin{abstract}

Previously identified integrable boundary states in ABJM theory are exclusively achiral. This paper presents the first chiral integrable boundary states in the $SU(4)$ alternating spin chain from the planar two-loop dilatation operator in the scalar sector. Utilizing a sufficient condition for the untwisted integrable condition, we identify specific two-site and four-site basis boundary states as chiral integrable states. Numerical evidence indicates that other basis states are unlikely to be chiral integrable. Furthermore, we compute the overlaps between these chiral integrable basis states and on-shell Bethe eigenstates. 
\end{abstract}

\tableofcontents

\section{Introduction}

Integrable boundary states, more precisely understood as integrable initial states in their physical interpretation, were first studied in the context of two-dimensional integrable quantum field theories with boundaries~\cite{Ghoshal:1993tm}. 
These states are characterized by the fact that they are annihilated by all odd conserved charges. In 2017, Piroli, Pozsgay, and Vernier proposed an analogous definition for integrable boundary states $\left| \mathcal{B} \right\rangle$ in lattice models \cite{Piroli:2017sei}:

\begin{align}
Q_{2k+1}\left | \mathcal{B} \right \rangle=0,k\ge 1, \label{defbq}
\end{align}
where the conserved charges $Q_{n+1},n\ge 1$ are  generated from the  transfer matrix $\tau\left ( u \right )$ via
\begin{align}
Q_{n+1}\propto \left.\frac{\partial ^n}{\partial u ^n} \ln \tau\left ( u \right )\right|_{u=u_{\ast}},
\end{align}
with a suitable choice of $u_{\ast}$ such that $Q_{n+1}$ can be expressed as a sum over operators with finite range. 
Since the odd conserved charges are generated by the transfer matrix, a sufficient condition for a boundary state to be integrable can be formulated directly in terms of transfer matrix $\tau(u)$:
\begin{align}
\Pi \tau\left ( u \right ) \Pi \left | \mathcal{B} \right \rangle=\tau\left ( u \right ) \left | \mathcal{B} \right \rangle, \label{IBS}
\end{align} 
with $\Pi$ being the reflection operator
\begin{align}
    \Pi \left |A_1 A_2 \cdots A_L\right \rangle
    =\left| A_L \cdots A_2 A_1 \right\rangle.
\end{align}

%which, in our derivation, turns out to be more tractable than the condition \eqref{defbq}.

Motivated by significant applications in both quench dynamics of spin chains and gauge/string duality, there has been considerable interest in deriving explicit overlap formulas between integrable boundary states and eigenstates (on-shell Bethe states) of the transfer matrix. These overlaps exhibit several universal features.
Certain selection rules emerge, determining when the overlap can be non-vanishing. One crucial  selection rule  demands  the Bethe roots characterizing a given eigenstate must exhibit a specific pairing structure. This is because integrable states preserve infinitely many conserved charges, as shown in~(\ref{defbq}). In higher-rank spin chains, Bethe roots can be paired in different ways \cite{Gombor:2020kgu}. An overlap is referred to as chiral if the pairing occurs among Bethe roots of the same type, and achiral if the pairing involves roots of different types. Furthermore, the non-vanishing overlaps involving integrable boundary states and eigenstates are relatively simple, compared to non-integrable cases, often admitting compact expressions in terms of ratios of Gaudin-like determinants multiplied by scalar factors which are functions of Bethe roots. 

In the context of the $AdS_{5}/CFT_{4}$ correspondence, both types of pairing structures have appeared in the literature, notably in studies of correlation functions in the presence of defects and three-point functions \cite{deLeeuw:2015hxa, Buhl-Mortensen:2015gfd, deLeeuw:2016umh, deLeeuw:2017dkd, Buhl-Mortensen:2017ind, DeLeeuw:2018cal, DeLeeuw:2019ohp, Kristjansen:2020mhn, Jiang:2019xdz, Jiang:2019zig, Kristjansen:2023ysz, Gombor:2024api, Kristjansen:2024map, Holguin:2025bfe, Chalabi:2025nbg, Coronado:2025xwk} . In contrast, within the $AdS_{4}/CFT_{3}$ framework, only the achiral pairing structure with pairing condition ${\mathbf{u} } = {-\mathbf{v} }$ and ${\mathbf{w} } = {-\mathbf{w} }$ imposed on the sets of Bethe roots $\mathbf{u}, \mathbf{w}, \mathbf{v}$ has been observed~\cite{Yang:2021hrl, Kristjansen:2021abc, Gombor:2022aqj, Yang:2022dlk, Jiang:2023cdm, Wu:2024uix, Bai:2024qtg}~\footnote{On the string theory side, the integrability of open strings ending on a $D4$-brane dual to a domain wall was revealed in \cite{Linardopoulos:2022wol}, based on earlier studies on dynamical reflection matrices~\cite{Linardopoulos:2021rfq}. See \cite{Linardopoulos:2025ypq} for a very nice review on this subject.} .

This paper aims to fill this gap by identifying special two-site and four-site states which are  chiral integrable states among the basis states. 
We also have numerical evidence which supports that 
the other basis states are not chiral integrable boundary states.

The planar two-loop dilatation operator acting on the scalar sector of ABJM theory~\cite{Aharony:2008ug} gives an integrable Hamiltonian of an alternating spin chain~\cite{Minahan:2008hf, Bak:2008cp}.  States on odd/even sites of this chain are in the fundamental/anti-fundamental representation. In this case, there are two transfer matrices $\tau(u)$, $\bar{\tau}(u)$ which commute for arbitrary spectral parameters
\begin{align}
[\tau(u), \tau(v)]=[\bar{\tau}(u), \bar{\tau}(v)]=[\tau(u), \bar{\tau}(v)]=0.
\end{align}
These transfer matrices generate two sets of conserved charges $Q_l, \bar{Q}_l$.
Two types of integrable boundary states are anticipated. They are chiral integrable states satisfying the untwisted integrable condition

\begin{align}\label{untwisted}
    \Pi \tau(u)\Pi \left |\mathcal{B} \right\rangle =\tau(u)\left |\mathcal{B} \right\rangle,
\end{align}
and
achiral integrable states satisfying the twisted integrable condition 
\begin{align}\label{twisted}
    \Pi \tau(u)\Pi \left |\mathcal{B} \right\rangle =\bar{\tau}(u)\left |\mathcal{B} \right\rangle.
\end{align}
In this paper, we focus on the chiral integrable states. Instead of dealing with integrable conditions in terms of conserved charges $Q_l$ and $\bar{Q}_l$, we found that taking \eqref{untwisted} as a starting point
turns out to be more tractable.

The remainder of this paper is organized as follows. In \autoref{section set up}, we review the fundamental setup of the $SU(4)$ alternating spin chain. Then \autoref{sec:Explicit calculation} is devoted to identifying chiral integrable boundary states through a detailed derivation. Overlap formulas between these boundary states and Bethe states are presented in \autoref{section overlap}. Finally, we give a conclusion in \autoref{section conclusion}. Several Bethe roots used in our numerical checks are listed in \autoref{bethe root}.

\section{ABJM spin chain}\label{section set up}

We begin by reviewing the $SU(4)$ alternating spin chain from ABJM theory.
Gauge invariance requires that single-trace operators be represented as states in an  alternating $SU(4)$ spin chain. States on odd sites of this chain transform in the fundamental representation of the R-symmetry group $SU(4)$, and those on even sites transform in the anti-fundamental representation. The planar two-loop dilatation operator acting on the scalar sector of ABJM theory can be described by the following spin chain Hamiltonian \cite{Minahan:2008hf, Bak:2008cp}
\begin{align}
    H = \frac{\lambda^2}{2} \sum_{l=1}^{2L} \left( 2 - 2P_{l, l+2} + P_{l, l+2}K_{l, l+1}+K_{l, l+1}P_{l, l+2} \right), \label{H}
\end{align}
where $P_{ij}$ and $K_{ij}$ are permutation and trace operators, respectively. Periodic boundary condition is understood.  
The orthonormal bases of the local Hilbert spaces are \( \left| A \right\rangle \) (\( A = 1, \cdots, 4 \)) for the fundamental representation sites and \( \left| \bar{A} \right\rangle \) (\( \bar{A} = \bar{1}, \cdots, \bar{4} \)) for the anti-fundamental representation ones. The permutation operator $P$ and trace operator act as
\begin{align}
P\left | A \right \rangle \otimes \left | B \right \rangle  &= \left | B \right \rangle \otimes  \left | A \right \rangle,\quad 
P\left | \bar{A}  \right \rangle \otimes \left | \bar{B}  \right \rangle  = \left | \bar{B}  \right \rangle \otimes  \left | \bar{A}  \right \rangle;\\
K\left | A \right \rangle \otimes \left | \bar{B} \right \rangle&=\delta_{AB} \sum_{C=1}^{4} \left | C \right \rangle \otimes \left | \bar{C}  \right \rangle, \quad 
K\left | \bar{A}  \right \rangle \otimes \left | B  \right \rangle  = \delta_{AB} \sum_{C =1}^{4} \left | \bar{C} \right \rangle \otimes \left | C \right \rangle.
\end{align}
It is also convenient to introduce the index notation for these operators. For example,
\begin{align}
I_{a_1a_2}^{b_1b_2}=\delta_{a_1}^{b_1}\delta_{a_2}^{b_2},\quad P_{a_1a_2}^{b_1b_2}=\delta_{a_1}^{b_2}\delta_{a_2}^{b_1},\quad K_{a_1\bar{a}_2}^{b_1\bar{b}_2}=\delta^{\bar{b}_1\bar{b}_2}\delta_{a_1a_2},\label{abAB}
\end{align}
%where $a,b$ is index of the auxiliary space and $A,B$ for physical space.
$P_{ij}$ and $K_{ij}$ satisfy the following relations which will be used extensively in subsequent derivations in \autoref{subsection pf}
\begin{align}
P_{ij}P_{ij}&=I_{ij},\label{pijpij}\\
P_{ij}P_{ik}&=P_{ik}P_{kj}=P_{jk}P_{ij},\label{pijpik}\\
K_{ij}K_{ij}&=4K_{ij},\label{kijkij}\\
K_{ij}K_{ik}&=K_{ij}P_{jk}=P_{jk}K_{ik},\label{kijkik}\\
\textbf{tr}_iP_{ij}&=I_j,\quad\textbf{tr}_iK_{ij}=I_j.\label{tr=i}
\end{align}
 
The $SU(4)$ alternating spin chain has four $R$ matrices,
\begin{align}
R_{0j}(u) &= u I + P_{0j}, \nonumber \\
R_{0\bar{j}}(u) &= -(u + 2) I + K_{0\bar{j}}, \nonumber \\
R_{\bar{0}j}(u) &= -(u + 2) I + K_{\bar{0}j}, \nonumber \\
R_{\bar{0}\bar{j}}(u) &= u I + P_{\bar{0}\bar{j}}, \label{R}
\end{align}
in which $0$ and $\bar{0}$ denote the auxiliary space in the fundamental and anti-fundamental representations of the R symmetry, respectively. $2L$ sites of the alternating spin chain are labeled as $1, \bar{1}, 2, \bar{2},\cdots, L,\bar{L}$. Based on these, two monodromy matrices are defined as
\begin{align}
    T_0(u) &= R_{01}(u) R_{0\bar{1}}(u) \cdots R_{0L}(u) R_{0\bar{L}}(u),\label{M0}\\
    T_{\bar{0}}(u) &= R_{\bar{0}1}(u) R_{\bar{0}\bar{1}}(u) \cdots R_{\bar{0}L}(u) R_{\bar{0}\bar{L}}(u).\label{M0b}
\end{align} 
The associated transfer matrices are obtained by tracing over the auxiliary space 
\begin{align}
\tau(u) &= \text{tr}_0 T_0(u), \quad \bar{\tau}(u) = \text{tr}_{\bar{0}} T_{\bar{0}}(u). \label{t}
\end{align}

A general on-shell Bethe state is characterized by three sets of Bethe roots, $\mathbf{u}$, $\mathbf{w}$ and $\mathbf{v}$, which satisfy a set of Bethe equations
\begin{align}
    \begin{gathered}\begin{aligned}1=\left(\frac{u_j+\frac{i}{2}}{u_j-\frac{i}{2}}\right)^L\prod_{\begin{array}{c}k=1\\k\neq j\end{array}}^{N_{\mathbf{u}}}S(u_j,u_k)\prod_{k=1}^{N_{\mathbf{w}}}\tilde{S}(u_j,w_k)\end{aligned}\\\begin{aligned}1=\prod_{\begin{array}{c}k=1\\k\neq j\end{array}}^{N_{\mathbf{w}}}S(w_j,w_k)\prod_{k=1}^{N_{\mathbf{u}}}\tilde{S}(w_j,u_k)\prod_{k=1}^{N_{\mathbf{v}}}\tilde{S}(w_j,v_k)\end{aligned}\\\begin{aligned}1=\left(\frac{v_j+\frac{i}{2}}{v_j-\frac{i}{2}}\right)^L\prod_{\begin{array}{c}k=1\\k\neq j\end{array}}^{N_\mathbf{v}}S(v_j,v_k)\prod_{k=1}^{N_\mathbf{w}}\tilde{S}(v_j,w_k),\end{aligned}\end{gathered}\label{bae}
\end{align}
where $N_u, N_w, N_v$ denote the number of rapidities in the sets $\mathbf{u} = \{u_1, \dots, u_{N_u}\}$, $\mathbf{w} = \{w_1, \dots, w_{N_w}\}$, $\mathbf{v} = \{v_1, \dots, v_{N_v}\}$, and
\begin{align}
S(u, v) &= \frac{u - v - i}{u - v + i}, \quad \tilde{S}(u, v) = \frac{u - v + \frac{i}{2}}{u - v - \frac{i}{2}}. \label{s}
\end{align}

Finally, the eigenvalues of the transfer matrices are expressed as
\begin{align}
\Lambda(u)=&
(-u-2)^L(u+1)^L\prod_{i=1}^{K_u}\frac{u-iu_i-1/2}{u-iu_i+1/2}\nonumber
\\
&+(-u)^L(u+1)^L\prod_{i=1}^{K_v}\frac{u-iv_i+5/2}{u-iv_i+3/2}\nonumber\\
&+(-u)^L(u+2)^L\prod_{i=1}^{K_u}\frac{u-iu_i+3/2}{u-iu_i+1/2}\prod_{i=1}^{K_w}\frac{u-iw_i}{u-iw_i+1}\nonumber\\
&+(-u)^L(u+2)^L\prod_{i=1}^{K_v}\frac{u-iv_i+1/2}{u-iv_i+3/2}\prod_{i=1}^{K_w}\frac{u-iw_i+2}{u-iw_i+1}, \label{lmd}
\end{align}
\begin{align}
    \bar{\Lambda}(u)=&(-u)^L(u+1)^L\prod_{i=1}^{K_u}\frac{u-iu_i+5/2}{u-iu_i+3/2}\nonumber\\
    &+(-u-2)^L(u+1)^L\prod_{i=1}^{K_v}\frac{u-iv_i-1/2}{u-iv_i+1/2}\nonumber\\
    &+(-u)^L(u+2)^L\prod_{i=1}^{K_u}\frac{u-iu_i+1/2}{u-iu_i+3/2}\prod_{i=1}^{K_w}\frac{u-iw_i+2}{u-iw_i+1}\nonumber\\
    &+(-u)^L(u+2)^L\prod_{i=1}^{K_v}\frac{u-iv_i+3/2}{u-iv_i+1/2}\prod_{i=1}^{K_w}\frac{u-iw_i}{u-iw_i+1}.\label{lmdb}
\end{align}
The corresponding Bethe states $\left | \bf{u}, \bf{v}, \bf{w}\right\rangle$ can be constructed using nested 
coordinate Bethe ansatz~\cite{Yang:2021hrl} where the vacuum was chosen to be $\left |1\bar{4}\right\rangle^{\otimes L}$.

\section{Chiral Integrable Boundary States} \label{sec:Explicit calculation}

By explicitly examining the expressions for $\Lambda(u) \eqref{lmd}$ and $\bar{\Lambda}(u) \eqref{lmdb}$, we observe that when $\left \{ \mathbf{u} \right \} =\left \{ -\mathbf{u} \right \} ,
\left \{ \mathbf{w} \right \} =\left \{ -\mathbf{w} \right \} , 
\left \{ \mathbf{v} \right \} =\left \{ -\mathbf{v} \right \} $, the relation $\Lambda (u)=\bar{\Lambda } (-u-2)$ holds. This motivates the expectation that the integrability condition
\begin{align}
    \tau (u) \left | \mathcal{B}  \right \rangle=\bar{\tau } (-u-2) \left | \mathcal{B}  \right \rangle \label{tbc}
\end{align}
leads to the chiral pairing structure
\begin{align}
    \left \{ \mathbf{u} \right \} =\left \{ -\mathbf{u} \right \} ,\left \{ \mathbf{w} \right \} =\left \{ -\mathbf{w} \right \} , \left \{ \mathbf{v} \right \} =\left \{ -\mathbf{v} \right \}
\end{align}
for the overlap $\left \langle \mathcal{B} | \mathbf{u}, \mathbf{w}, \mathbf{v} \right \rangle $ to be non-vanishing.

In fact, $\eqref{tbc}$ can be derived from the untwisted integrable condition \cite{Gombor:2020kgu} 
\begin{align}
\Pi \tau (u)\Pi \left | \mathcal{B}  \right \rangle & = \tau (u) \left | \mathcal{B}  \right \rangle, 
\end{align}
where $\Pi$ denotes the spatial reflection operator defined by 
\begin{align}
\Pi \left| B^1 \bar{B}_1 B^2 \bar{B}_2 \cdots B^L \bar{B}_L \right\rangle=
\left| \bar{B}_L B^L \cdots \bar{B}_2 B^2 \bar{B}_1 B^1 \right\rangle.
\end{align}
The transformation $\Pi \tau(u) \Pi$ reverses the order of R-matrix products
\begin{align}
\Pi \tau(u) \Pi = \text{tr}_0 R_{0\bar{L}}(u) R_{0L}(u) \cdots R_{0\bar{1}}(u) R_{01}(u).
\end{align}
Using crossing symmetry relations for R-matrices
\begin{align}
R_{0\bar{j}}^{t_a}(u)=R_{\bar{0}\bar{j}}(-u-2), R_{0j}^{t_a}(u)=R_{\bar{0}j}(-u-2),
\end{align}
where $t_a$ denotes transposition with respect to auxiliary space, we deduce that
\begin{align}
\Pi \tau(u) \Pi 
&= \text{tr}_0
\left ( R_{\bar{0}1}(-u-2) R_{\bar{0}\bar{1}}(-u-2) \cdots 
R_{\bar{0}L}(-u-2) R_{\bar{0}\bar{L}}(-u-2)\right )^{t_a} \\ \nonumber 
&=\bar{\tau}\left ( -u-2 \right ).
\end{align}
Thus gives rise to $\eqref{tbc}$.

We now aim to identify the states that satisfy the integrability condition $\eqref{tbc}$. In general, such a state is a specific and intricate linear combination of the orthonormal basis states spanning the full Hilbert space of the spin chain. For simplicity, however, we restrict our attention to basis states $\left| B \right\rangle = \left| B^1 \bar{B}_1 B^2 \bar{B}_2 \cdots B^L \bar{B}_L \right\rangle$, with $B^i = 1,2,3,4,\ i = 1,\cdots,L$ and $\bar{B}_j = \bar{1},\bar{2},\bar{3},\bar{4},\ j = 1,\cdots,L$, and ask which of these are integrable. 

To further facilitate the analysis, we expand the transfer matrix in a specific form and derive a sufficient condition for the integrability condition $\eqref{tbc}$, as done for achiral integrable boundary states in the ABJM spin chain~\cite{Yang:2022dlk, Wu:2024uix}. Starting from $\eqref{t}$, we write
\begin{align}
\tau(u)
&=\sum_{m,n=0}^{L} u^{L-m}\left ( -(u+2) \right ) ^{L-n}\mathcal{O}_{m,n}\ ,\label{ext}
\end{align}
where, omit all identity operators $I$,
\begin{align}
\mathcal{O}_{m,n} =\sum_{\substack{i_1<\cdots <i_m\\ \bar{j}_1<\cdots <\bar{j}_n}}\text{tr}_0\left ( 
P_{0i_1}\cdots K_{0\bar{j}_1 }\cdots K_{0\bar{j}_n}\cdots P_{0i_m} \right ) ,\label{omn}
\end{align}
where $i_k,j_l\in \left \{ 1, \cdots, L \right \}$, or any positive integer with periodic boundary condition understood. Note that here we always have $m$ permutation operators $P$ and $n$ trace operators $K$ in $\mathcal{O}_{m,n}$ and are ordered according to the second index in their subscript, which corresponds to their position in the spin chain.

Similarly, starting from \eqref{t} we get the following expansion for $\bar{\tau}\left ( -u-2 \right )$
\begin{align}
\bar{\tau}\left ( -u-2 \right )&=\sum_{m,n=0}^{L} u^{L-m}\left ( -(u+2) \right ) ^{L-n}\overline{\mathcal{O}}_{m,n}\ ,\label{extb} \\
\overline{\mathcal{O}}_{m,n} &=\sum_{\substack{i_1<\cdots <i_m\\ \bar{j}_1<\cdots <\bar{j}_n}}\text{tr}_{\bar{0}}\left ( 
K_{\bar{0} i_1}\cdots P_{\bar{0}\bar{j}_1 }\cdots P_{\bar{0}\bar{j}_n}\cdots K_{\bar{0}i_m} \right ),\label{obmn}
\end{align}
where $\overline{\mathcal{O}}_{m,n}$ contains $m$ trace operators $K$ and $n$ permutation operators $P$. 

Once we explicitly specify the number of $P$ and $K$ operators included in the above $\mathcal{O}_{m,n}$ and $\overline{\mathcal{O}}_{m,n}$ operators, the above expansions $\eqref{ext}$, $\eqref{extb}$ become unique, with a total of $(L+1)^2$ terms. Using the expansion for transfer matrices $\eqref{ext}$, $\eqref{extb}$, we can write down the sufficient condition for the integrable condition $\eqref{tbc}$
\begin{align}
    \mathcal{O}_{m,n}\left | \mathcal{B} \right \rangle = \overline{\mathcal{O}}_{m,n}\left | \mathcal{B} \right \rangle,\ m,n=0,\cdots,L.\label{omn=obmn}
\end{align}

Note that there also exists an alternative expansion with a total of $(2L+1)$ terms
\begin{align}
\tau(u)&=\sum_{s=0}^{2L} u^{s}\hat{\mathcal{O}}_s,\\
\bar{\tau}(-u-2)&=\sum_{s=0}^{2L} u^{s}\hat{\overline{\mathcal{O}}}_s,
\end{align}
where the operators $\hat{\mathcal{O}}_s$ and $\hat{\overline{\mathcal{O}}}_s$ are generally more involved, being linear combinations of those defined in $\eqref{omn}$ and $\eqref{obmn}$. In fact, one equivalent condition of $\eqref{tbc}$ is given by
\begin{align}
\hat{\mathcal{O} }_s\left | \mathcal{B}\right \rangle  = \hat{\overline{\mathcal{O}}}_s \left | \mathcal{B}\right \rangle, s & = 0, \cdots, 2L.\label{ohs=ohbs}
\end{align}
However, due to the complexity of $\hat{\mathcal{O}}_s$ and $\hat{\overline{\mathcal{O}}}_s$, we prefer to work with the simpler form in $\eqref{omn=obmn}$.

Therefore, the problem reduces to determining which  basis states 
\begin{equation}\left | B \right \rangle =\left | B^1 \bar{B}_1 B^2 \bar{B}_2 \cdots B^L \bar{B}_L \right \rangle, B^i=1,2,3,4, \bar{B}_j=\bar{1},\bar{2},\bar{3},\bar{4},
\end{equation}
satisfy the integrability condition
\begin{align}
    \mathcal{O}_{m,n}\left | B \right \rangle = \overline{\mathcal{O}}_{m,n}\left | B \right \rangle,\ m,n=0,\cdots,L.\label{omnb=obmnb}
\end{align}

In the following, we first display the final results in \autoref{subsection chiral} and then leave the detailed derivation in \autoref{subsection pf}.

\subsection{Chiral Integrable Boundary States}\label{subsection chiral}

Solving the equations $\eqref{omnb=obmnb}$, we identify four types of integrable basis states $\left| B \right\rangle$
\begin{align}
\left | B \right \rangle = \begin{cases}
\left |A\bar{B}\cdots A\bar{B} \right \rangle, A\ne B  & Any\ L \\
\left |A\bar{B} A\bar{D}\cdots A\bar{B} A\bar{D} \right \rangle, A, B, D\ are\ all\ mutually\ distinct  & L\ even\\
\left |A\bar{B} C\bar{B}\cdots A\bar{B} C\bar{B} \right \rangle, A, B, C\ are\ all\ mutually\ distinct  & L\ even\\
\left |A\bar{B} C\bar{D}\cdots A\bar{B} C\bar{D} \right \rangle, A, B, C, D\ are\ all\ mutually\ distinct  & L\ even\end{cases}
\label{4types}
\end{align}
where $A,C\in \{1,2,3,4\}$,$\bar{B},\bar{D}\in \{\bar{1},\bar{2},\bar{3},\bar{4}\}$. These results show that two-site basis states are integrable for arbitrary chain length $L$. These two-site basis states correspond to the $1/3$-BPS chiral primary operators in ABJM theory. Moreover, for even $L$, three additional classes of four-site states are also integrable.

We have explicitly verified that for $L=2$, the conditions $\eqref{omn=obmn}$ and $\eqref{ohs=ohbs}$ select the same set of integrable states. For higher $L$, we employ explicit unpaired Bethe root solutions to numerically generate the non-integrable basis states—those that appear in a Bethe eigenstate characterized by unpaired roots with non-vanishing coefficients. We find $460$ non-integrable basis states for $L=3$, 5146 for $L=4$, and 10240 for $L=5$. These basis states are all distinct from those listed in \eqref{4types}. These numerical results support the conclusion that the chiral integrable states among the basis states are precisely those given in \eqref{4types}. 

\subsection{Proof}\label{subsection pf}

To systematically solve the equations $\eqref{omnb=obmnb}$, we organize our analysis according to the values of $m$ and $n$ in the operators $\mathcal{O}_{m,n}$ and $\overline{\mathcal{O}}_{m,n}$. The overall strategy can be sketched as follows. First, we consider the case $n = 0$ or $m = 0$, where the resulting integrable boundary basis states exhibit specific periodic structures. Next, we examine the cases $n = 1$ or $m = 1$, in which the operators contain exactly one trace or one permutation operator. By performing explicit computations, additional constraints on the boundary states are obtained. Finally, for higher-order cases with $n \ge 2$ or $m \ge 2$, it is observed that if $\left | B \right \rangle$ exhibits the structural patterns identified in the previous steps, then the operator actions vanish automatically, ensuring that the integrability condition is satisfied without further restrictions. 

In the following, we present the detailed derivations corresponding to each of the cases outlined above.

\subsubsection{\texorpdfstring{$n=0$ or $m=0$}{n=0 or m=0}}

We begin by analyzing the operator structures in the simplest cases where either $n = 0$ or $m = 0$. Upon taking the trace over the auxiliary space $0$ explicitly, these operators contain only permutation operators and act either on the fundamental or anti-fundamental sites of the alternating spin chain. Concretely, they take the following forms:

\begin{align}
\mathcal{O}_{m,0} = \sum_{i_1<\cdots<i_m}P_{i_mi_1}P_{i_mi_2}\cdots P_{i_mi_{m-1}}, \label{om0}
\end{align}
where $P_{i_mi_1}P_{i_mi_2}\cdots P_{i_mi_{m-1}}$ cyclically shifts the states at positions $i_1,\cdots,i_m$ one site to the left, within this subset of sites.

\begin{align}
\overline{\mathcal{O}}_{m,0} = \sum_{i_1<\cdots<i_m}P_{i_1i_2}P_{i_2i_3}\cdots P_{i_{m-1}i_m}, \label{obm0}
\end{align}
where $P_{i_1i_2}P_{i_2i_3}\cdots P_{i_{m-1}i_m}$ instead shifts the same set of states one site to the right within the positions $i_1,\cdots,i_m$.

Likewise, the operators acting purely on the anti-fundamental sites take the form:

\begin{align}
\mathcal{O}_{0,n}=\sum_{\bar{j}_1<\cdots < \bar{j}_n}P_{\bar{j}_1 \bar{j}_2}P_{\bar{j}_2 \bar{j}_3}\cdots P_{\bar{j}_{n-1} \bar{j}_n}, \label{o0n} 
\end{align}
where $P_{\bar{j}_1 \bar{j}_2}P_{\bar{j}_2 \bar{j}_3}\cdots P_{\bar{j}_{n-1} \bar{j}_n}$ shifts the states at positions $\bar{j}_1,\cdots,\bar{j}_n$ one site to the right within the selected subset.

\begin{align}
\overline{\mathcal{O}}_{0,n}=\sum_{\bar{j}_1<\cdots < \bar{j}_n}P_{\bar{j}_n \bar{j}_1}P_{\bar{j}_n \bar{j}_2}\cdots P_{\bar{j}_n \bar{j}_{n-1} }, \label{ob0n} 
\end{align}
where $P_{\bar{j}_n \bar{j}_1}P_{\bar{j}_n \bar{j}_2}\cdots P_{\bar{j}_n \bar{j}_{n-1} }$ performs a one-site leftward shift on the same set of positions $\bar{j}_1,\cdots,\bar{j}_n$.

\paragraph{\texorpdfstring{$m=L$ or $n=L$}{m=L or n=L}} \ 
\newline
For the case $m=L,n=0$, the integrability condition takes the form
\begin{align}
\mathcal{O}_{L,0}\left | B \right \rangle &=\overline{\mathcal{O}}_{L,0}\left | B \right \rangle,\\
P_{L1} P_{L2}\cdots P_{L\left ( L-1 \right ) }\left | B \right \rangle&=P_{12} P_{23}\cdots P_{\left ( L-1 \right )L }\left | B \right \rangle.
\end{align}
where $\left | B \right \rangle =\left | B^1 \bar{B}_1 B^2 \bar{B}_2 \cdots B^L \bar{B}_L \right \rangle, B^i=1,2,3,4,\ i=1,\cdots,L,\ \bar{B}_j=\bar{1},\bar{2},\bar{3},\bar{4},\ j=1,\cdots,L$.

For $L$ odd, it follows from the equality of operator actions that all $B^i,i=1,\cdots,L$ must be identical. We denote this common value by $A\in \{1,2,3,4\}$. While for $L$ even, we find that $B^i$ alternates between two values on odd and even sites, i.e., $B^{2k-1} = A$ and $B^{2k} = C$, with $A, C$ $\in \{1,2,3,4\}$. We get similar results for the anti-fundamental sites of the chain from $m=L,n=0$ case. In summary, at this stage we have
\begin{align}
    \left | B \right \rangle=\begin{cases}
\left | A\bar{B}A\bar{B}\cdots A\bar{B}  \right \rangle   & \text{ if } L\ odd \\
\left | A\bar{B}C\bar{D}\cdots A\bar{B} C\bar{D}  \right \rangle  & \text{ if } L\ even
\end{cases} \label{bl0}
\end{align}
where $A,C\in \{1,2,3,4\}$,$\bar{B},\bar{D}\in \{\bar{1},\bar{2},\bar{3},\bar{4}\}$.

\paragraph{\texorpdfstring{$m=0,1,2$ or $n=0,1,2$}{m=0,1,2 or n=0,1,2}} \ 
\newline
It is clear that both $\mathcal{O}_{0,0}$ and $\overline{\mathcal{O}}_{0,0}$ consist solely of the identity operator. Consequently, the condition $\mathcal{O}_{0,0}\left|B\right\rangle = \overline{\mathcal{O}}_{0,0}\left|B\right\rangle$ is trivially satisfied for any state.

A straightforward computation further shows that
\begin{align}
\mathcal{O}_{1,0}=L\, I,\ \overline{\mathcal{O}}_{1,0}=L\, I,\\
\mathcal{O}_{0,1}=L\, I,\ \overline{\mathcal{O}}_{0,1}=L\,I.
\end{align}
Thus
\begin{align}
    \mathcal{O}_{1,0}\left |B\right \rangle=\overline{\mathcal{O}}_{1,0}\left |B\right \rangle, \ 
    \mathcal{O}_{0,1}\left |B\right \rangle=\overline{\mathcal{O}}_{0,1}\left |B\right \rangle
\end{align}
are again trivially satisfied for arbitrary boundary states.

In the case of $m=2$ or $n=2$, 
\begin{align}
\mathcal{O}_{2,0}=\sum_{i_1<i_2}P_{i_2i_1},\ \overline{\mathcal{O}}_{2,0}=\sum_{i_1<i_2}P_{i_1i_2},\\
\mathcal{O}_{0,2}=\sum_{\bar{j}_1<\bar{j}_2}P_{\bar{j}_2\bar{j}_1},\ \overline{\mathcal{O}}_{0,2}=\sum_{\bar{j}_1<\bar{j}_2}P_{\bar{j}_1\bar{j}_2}.
\end{align}
When acting on an arbitrary basis state $\left |B\right \rangle$, the two sides of each equation yield the same result.

In conclusion, all these operators impose no additional constraints on the boundary state for the integrability condition to hold.

\paragraph{\texorpdfstring{$3\le m<L$ or $3\le n<L$}{3≤m<L or 3≤n<L}} \ 
\newline
For L odd, clearly $\mathcal{O}_{m,0}$, $\overline{\mathcal{O}}_{m,0}$, $\mathcal{O}_{0,n}$ and $\overline{\mathcal{O}}_{0,n}$ act on boundary state $\left |B\right \rangle \eqref{bl0}$ gives $\left |B\right \rangle$ itself.

For L even, we conjecture that the equalities
\begin{align}
    \mathcal{O}_{m,0}\left |B\right \rangle=\overline{\mathcal{O}}_{m,0}\left |B\right \rangle, \ 
    \mathcal{O}_{0,n}\left |B\right \rangle=\overline{\mathcal{O}}_{0,n}\left |B\right \rangle
\end{align}
also hold automatically for boundary states $\left |B\right \rangle$ of the form given in $\eqref{bl0}$. 
Although we do not yet have a rigorous proof of this statement, we have explicitly verified its validity for system sizes up to $L\le 26$. This gives us strong confidence that this statement is correct.

\subsubsection{\texorpdfstring{$n=1$ or $m=1$}{n=1 or m=1}}

Evaluating the trace explicitly in the operators with $n=1$ or $m=1$ yields
\begin{align}
\mathcal{O}_{m,1}
&=
\sum_{\substack{i_1<\cdots<i_m\\\bar{j}_1\left (i_{k-1}\le j_1<i_k \right )}}
\text{tr}_0\left ( P_{0i_1}\cdots P_{0i_{k-1}}K_{0\bar{j}_1}P_{0i_k}\cdots P_{0i_m} \right )\\ \nonumber
&=
\sum_{\substack{i_1<\cdots<i_m\\\bar{j}_1\left (i_{k-1}\le j_1<i_k \right )}}
\text{tr}_0\left ( K_{0\bar{j}_1}P_{0i_k}\cdots P_{0i_m}\ P_{0i_1}\cdots P_{0i_{k-1}} \right )\\ \nonumber
&=
\sum_{\substack{i_1<\cdots<i_m\\\bar{j}_1\left (i_{k-1}\le j_1<i_k \right )}}
\text{tr}_0\left ( K_{0\bar{j}_1} P_{0i_{k-1}} \right )
\left ( P_{i_{k-1}i_k}\cdots P_{i_{k-1}i_m}P_{i_{k-1}i_1}\cdots P_{i_{k-1}i_{k-2}} \right )\\ \nonumber
&=
\sum_{\substack{i_1<\cdots <i_m \\ \bar{j}_1\left ( i_{k-1}\le j_1<i_k \right ) }}
K_{i_{k-1}\bar{j}_1}\left ( P_{i_{k-1} i_{k}}\cdots P_{i_{k-1} i_{m}}P_{i_{k-1} i_{1}}\cdots P_{i_{k-1} i_{k-2}} \right ). 
\end{align}
We omit the details for the rest derivation as the remaining cases can be derived in a similar manner.
\begin{align}
\overline{\mathcal{O}}_{m,1}=
\sum_{\substack{i_1<\cdots <i_m \\ \bar{j}_1\left ( i_{k-1}\le j_1<i_k \right ) }}
K_{\bar{j}_1i_k}\left ( P_{i_ki_{k+1}}\cdots P_{i_{m-1}i_{m}}P_{i_mi_{1}}\cdots P_{i_{k-2}i_{k-1}} \right ),
\end{align}
\begin{align}
\mathcal{O}_{1,n}=\sum_{\substack{i_1 \left ( j_{k-1}<i_1\le j_k \right ) \\ \bar{j}_1<\cdots <\bar{j}_n}} 
K_{i_1 \bar{j}_k}\left ( P_{\bar{j}_{k} \bar{j}_{k+1}}\cdots P_{\bar{j}_{n} \bar{j}_{1}}P_{\bar{j}_{1} \bar{j}_{2}}\cdots P_{\bar{j}_{k-2} \bar{j}_{k-1}}  \right ), 
\end{align}
\begin{align}
\overline{\mathcal{O}}_{1,n}=\sum_{\substack{i_1 \left ( j_{k-1}<i_1\le j_k \right ) \\ \bar{j}_1<\cdots <\bar{j}_n}}
K_{\bar{j}_{k-1}i_{1}}\left ( P_{\bar{j}_{k-1}\bar{j}_{k}}\cdots P_{\bar{j}_{k-1}\bar{j}_{n}}P_{\bar{j}_{k-1}\bar{j}_{1}}\cdots P_{\bar{j}_{k-1}\bar{j}_{k-2}} \right ). 
\end{align}

\paragraph{\texorpdfstring{$m=L, n=1$}{m=L,n=1}} \ 
\newline
In this case, the operators are given by
\begin{align}
\mathcal{O}_{L,1}&= \sum_{\bar{j}_1=\bar{1}}^{\bar{L}}K_{j_{1}\bar{j}_1}\left ( P_{j_1\left ( j_1+1\right)}\cdots P_{j_1L}P_{j_11}\cdots P_{j_1\left ( j_1-1\right)}\right).\label{ol1}\\
\overline{\mathcal{O}}_{L,1}&=\sum_{\bar{j}_1=\bar{1}}^{\bar{L}}K_{\bar{j}_1(j_1+1)}\left ( P_{\left ( j_1+1\right)\left ( j_1+2\right)}\cdots P_{(L-1)L}P_{L1}\cdots P_{\left ( j_1-1\right)j_1}\right).\label{obl1}
\end{align}
We aim to study further constraints from  the condition
\begin{align}
\mathcal{O}_{L,1} \left |B\right \rangle = \overline{\mathcal{O}}_{L,1} \left |B\right \rangle,
\end{align}
for boundary states $\left |B\right \rangle$ of the form given in $\eqref{bl0}$. The analysis is carried out separately for odd and even L.

When $L$ is odd, the permutation operators contained in $\eqref{ol1}$ and $\eqref{obl1}$ act trivially on the state $\left |B\right \rangle$, leaving it invariant. Therefore, 
\begin{align}
\mathcal{O}_{L,1}\left |B\right \rangle
&=\sum_{\bar{j}_1=\bar{1}}^{\bar{L}}K_{j_1\bar{j}_1} \left |B\right \rangle \nonumber \\
&=\delta _{AB}\sum_{\bar{j}_1=\bar{1}}^{\bar{L}}\sum_{C=1}^{4} \left | A\bar{B}\cdots \underset{j_1}{C}\bar{C} \cdots A\bar{B}\right \rangle, \label{ol1b} \\
\overline{\mathcal{O}}_{L,1}\left |B\right \rangle
&=\sum_{\bar{j}_1=\bar{1}}^{\bar{L}}K_{\bar{j}_1(j_1+1)} \left |B\right \rangle \nonumber \\
&=\delta _{AB}\sum_{\bar{j}_1=\bar{1}}^{\bar{L}}\sum_{C=1}^{4} \left | A\bar{B}\cdots A\bar{C}  \underset{j_1+1}{C}\bar{B} \cdots A\bar{B}\right \rangle. \label{obl1b}
\end{align}
By comparing $\eqref{ol1b}$ and $\eqref{obl1b}$, we find that the equality holds if and only if $\delta _{AB}=0$.

For even $L$, the boundary state takes the form $\left | B \right \rangle=\left | A\bar{B}C\bar{D}\cdots A\bar{B} C\bar{D}  \right \rangle$. The action of the shift operators gives
\begin{align}
\left ( P_{j_1\left ( j_1+1 \right ) }\cdots P_{j_1L}P_{j_11}\cdots P_{j_1\left ( j_1-1 \right ) }\right )\left | B \right \rangle
&=\left | C\bar{B}A\bar{D}\cdots C\bar{B}A\bar{D} \right \rangle, \\
\left ( P_{\left ( j_1+1 \right )\left ( j_1+2 \right ) }\cdots P_{\left ( L-1 \right )L }P_{L1}\cdots P_{\left ( j_1-1 \right )j_1 } \right )\left| B \right\rangle
&=\left | C\bar{B}A\bar{D}\cdots C\bar{B}A\bar{D} \right \rangle.
\end{align}
Hence,
\begin{align}
\mathcal{O}_{L,1}\left |B\right \rangle
&=
\sum_{\bar{j}_1=\bar{1}}^{\bar{L}}
K_{j_1\bar{j}_1}\left | C\bar{B}A\bar{D}\cdots C\bar{B}A\bar{D} \right \rangle\\ \nonumber
&=
\sum_{\bar{i}_1=\bar{1}}^{\overline{L/2}} \sum_{E=1}^{4} 
\left (\delta_{CB}\left |\cdots \underset{\text{site }\left (2i_1-1\right )}{E}\bar{E}A\bar{D}\cdots\right \rangle 
+\delta _{AD}\left |\cdots C\bar{B}\underset{\text{site }2i_1}{E}\bar{E}\cdots\right \rangle \right ),\\
\overline{\mathcal{O}}_{L,1}\left |B\right \rangle
&=
\sum_{\bar{j}_1=\bar{1}}^{\bar{L}}
K_{\bar{j}_1\left ( j_1+1 \right ) }\left | C\bar{B}A\bar{D}\cdots C\bar{B}A\bar{D} \right \rangle\\ \nonumber
&=
\sum_{\bar{i}_1=\bar{1}}^{\overline{L/2}} \sum_{E=1}^{4} 
\left (\delta_{BA}\left |\cdots C\bar{E}\underset{\text{site }2i_1}{E}\bar{D}\cdots\right \rangle 
+\delta _{DC}\left |\cdots C\bar{B}A\bar{E}\underset{\text{site }\left ( 2i_1+1 \right ) }{E}\bar{B}A\bar{D}\cdots\right \rangle \right ).\\
\end{align}
For these two expressions to be equal, we must have $\delta _{AB}=0, \delta _{AD}=0, \delta _{CB}=0, \delta _{CD}=0$.

In summary, the above analysis leads to the following four classes of admissible boundary states $\left | B \right \rangle$
\begin{align}
\left | B \right \rangle = \begin{cases}
\left |A\bar{B}\cdots A\bar{B} \right \rangle, A\ne B  & Any\ L \\
\left |A\bar{B} A\bar{D}\cdots A\bar{B} A\bar{D} \right \rangle, A, B, D\ are\ all\ mutually\ distinct  & L\ even\\
\left |A\bar{B} C\bar{B}\cdots A\bar{B} C\bar{B} \right \rangle, A, B, C\ are\ all\ mutually\ distinct  & L\ even\\
\left |A\bar{B} C\bar{D}\cdots A\bar{B} C\bar{D} \right \rangle, A, B, C, D\ are\ all\ mutually\ distinct  & L\ even\end{cases}\label{bl1}
\end{align}
where $A,C\in \{1,2,3,4\}$,$\bar{B},\bar{D}\in \{\bar{1},\bar{2},\bar{3},\bar{4}\}$.

\paragraph{\texorpdfstring{$m<L$ or $n\le L$}{m<L or n≤L}} \ 
\newline
Note that the solutions given in $\eqref{bl1}$ ensure that the odd and even sites of the entire chain take different values. Each term in the operators $\mathcal{O}_{m,1}$, $\overline{\mathcal{O}}_{m,1}$,$\mathcal{O}_{1,n}$, $\overline{\mathcal{O}}_{1,n}$ contains exactly one trace operator acting on both an odd and an even site. As a result, their action on the states in $\eqref{bl1}$ vanishes. Therefore, we conclude that the solutions in $\eqref{bl1}$ satisfy these integrability conditions automatically.

\subsubsection{\texorpdfstring{$n\ge 2$ or $m\ge 2$}{n≥2 or m≥2}}

As the values of $m$ and $n$ increase, the operators $\mathcal{O}$ and $\overline{\mathcal{O}}$ contain more trace operators. Due to the fact that the states in $\eqref{bl1}$ assign different values to even and odd sites, the action of these operators on such states continues to vanish. Consequently, the integrability condition is automatically satisfied for all relevant cases. This concludes our proof, and we identify the states in $\eqref{bl1}$ as the integrable boundary states of interest.

\section{Overlap $\left \langle B|\mathbf{u}, \mathbf{w}, \mathbf{v} \right \rangle$} \label{section overlap}

Next, we aim to derive the explicit formula for the overlap between a boundary state $\left| B \right\rangle$, which takes one of the forms listed in equation $\eqref{4types}$~\footnote{We exclude the trivial case when $\left |B \right \rangle$  is the vacuum state $ \left | 1\bar{4}\right \rangle^{\otimes L}$.}, and an on-shell Bethe state $\left| \mathbf{u}, \mathbf{w}, \mathbf{v} \right\rangle$. Our convention for the Hermitian conjugate of a state is as follows:
\begin{align}
\left | B \right \rangle =\left | B^1 \bar{B}_1 B^2 \bar{B}_2 \cdots B^L \bar{B}_L \right \rangle,\\
\left\langle B \right| =\left\langle B^1 \bar{B}_1 B^2 \bar{B}_2 \cdots B^L \bar{B}_L \right|.
\end{align}

For the $\left \langle B | \mathbf{u}, \mathbf{w}, \mathbf{v} \right \rangle$ to be non-vanishing, two selection rules must be satisfied.

First, $\left |B\right\rangle$  must appear in the Bethe eigenstate $\left | \mathbf{u}, \mathbf{w}, \mathbf{v} \right \rangle$ with non-vanishing coefficient. This gives constraints on  the quantum numbers $N_u$, $N_w$, and $N_v$ characterizing the Bethe states for given $\left |  B \right 
\rangle$. This can be understood easily  from the construction of the Bethe states via nested coordinate Bethe ansatz~\cite{Yang:2021hrl}.
Similar selection rule was found in~\cite{Yang:2021hrl, Jiang:2023cdm, Wu:2024uix}
and this rule is not related to the integrability nature of the state.

The second selection rule is: the rapidities $\{\mathbf{u}\}$, $\{\mathbf{w}\}$, and $\{\mathbf{v}\}$ specifying the on-shell Bethe state must exhibit a pairing structure, i.e., $\{\mathbf{u}\} = \{-\mathbf{u}\}$, $\{\mathbf{w}\} = \{-\mathbf{w}\}$, and $\{\mathbf{v}\} = \{-\mathbf{v}\}$,  for the chiral integrable boundary states as discussed at the beginning of \autoref{sec:Explicit calculation}.

Practically, we notice that  the rational Q-system~\cite{Marboe:2016yyn, Gu:2022dac} used to efficiently solve the Bethe ansatz equations can be used only when the quantum numbers satisfy the constraints $N_u, N_v \le \frac{1}{2}(L + N_w)$ and $N_w \le \frac{1}{2}(N_u + N_v)$. Taking these conditions into account, eight specific chiral boundary states are ``on the right side of the equator,'' as listed in \autoref{eight}. The remaining boundary states, characterized by quantum numbers ``on the wrong side of the equator''~\cite{Pronko:1998xa}, can be accessed through the bosonic duality procedure \cite{Gromov:2007ky}. As shown in \autoref{eight}, the eight chiral boundary states give rise to non-vanishing overlaps exclusively with Bethe states belonging to the $SU(2)$ or $SU(2) \times SU(2)$ sector. These overlaps are essentially governed by the well-established formulas for the overlap between the N\'eel state and Bethe eigenstates in the XXX spin chain, and have been extensively studied in the literature.

\begin{table}[!ht]
    \centering
    \begin{tabular}{|c|c|c|c|}
    \hline
        No. & $\left \langle B \right |$ & $\left ( N_u,N_w,N_v \right )$ & Sector \\ \hline
        1 & $\left \langle 2\bar{4}1\bar{4} \right | ^{\otimes \frac{L}{2} }$ & \multirow{2}{*}{$\left ( \frac{L}{2}, 0, 0 \right ) $} & \multirow{4}{*}{$SU(2)$} \\ \cline{1-2}
        2 & $\left \langle 1\bar{4}2\bar{4} \right | ^{\otimes \frac{L}{2} }$ & ~ & ~ \\ \cline{1-3}
        3 & $\left \langle 1\bar{3}1\bar{4} \right | ^{\otimes \frac{L}{2} }$ & \multirow{2}{*}{$\left ( 0, 0, \frac{L}{2}\right ) $} & ~ \\ \cline{1-2}
        4 & $\left \langle 1\bar{4}1\bar{3} \right | ^{\otimes \frac{L}{2} }$ & ~ & ~ \\ \hline
        5 & $\left \langle 2\bar{3}1\bar{4} \right | ^{\otimes \frac{L}{2} }$ & \multirow{4}{*}{$\left ( \frac{L}{2}, 0, \frac{L}{2} \right ) $} & \multirow{4}{*}{$SU(2) \times SU(2)$} \\ \cline{1-2}
        6 & $\left \langle 2\bar{4}1\bar{3} \right | ^{\otimes \frac{L}{2} }$ & ~ & ~ \\ \cline{1-2}
        7 & $\left \langle 1\bar{3}2\bar{4} \right | ^{\otimes \frac{L}{2} }$ & ~ & ~ \\ \cline{1-2}
        8 & $\left \langle 1\bar{4}2\bar{3} \right | ^{\otimes \frac{L}{2} }$ & ~ & ~ \\ \hline
    \end{tabular}
    \caption{Eight chiral integrable boundary states with quantum numbers compatible with Q-system. }
    \label{eight}
\end{table}

The remainder of this section is structured as follows. In \autoref{subsection wavefunction}, we review the conventions for Bethe eigenstate wavefunctions, the definition of the Gaudin determinant, and the expression for Bethe state norms in terms of the Gaudin determinant. In \autoref{subsection overlap}, we collect known overlap formulas relevant to our discussion, including results mainly from \cite{brockmann2014neel} and \cite{Jiang:2020sdw}. Using these results, we compute $\left \langle  B| \mathbf{u}, \mathbf{w}, \mathbf{v} \right\rangle$ for the boundary states in~\autoref{eight}. Also, the specific solutions of Bethe roots used in our numerical checks are presented in \autoref{bethe root}.

\subsection{Conventions for wavefunctions, Gaudin determinant and norms of Bethe states}\label{subsection wavefunction}

Our convention for wavefunction is in accordance with \cite{Yang:2021hrl}. Some useful functions are: 

\begin{align}
l_{u}: = e^{i p(u)} = \frac{u+\frac{i}{2}}{u-\frac{i}{2}},
\end{align}
\begin{align}
f(u,v)=\frac{u-v-i}{u-v} ,
\end{align}
\begin{align}
S(u,v)=\frac{u-v-i}{u-v+i} =\frac{f(u,v)}{f(v,u)}.
\end{align}

Gaudin determinant appears in various norm and  overlap formulas in spin chain systems. For parity-invariant Bethe roots, Gaudin determinant can be factorized into a product of two parts which depend on ``positive'' roots and ``negative'' ones respectively. Here we display explicit expression for Gaudin determinant of XXX spin chain.

For quantum numbers $(N_u, N_w, N_v) = \left(\frac{L}{2}, 0, 0\right)$, Bethe equations $\eqref{bae}$ of $SU(4)$ alternating spin chain reduced to

\begin{align}
1=&\left (\frac{u_j+\frac{i}{2}}{u_j-\frac{i}{2}}\right )^{L}
\prod_{\substack{k=1\\k\ne j}}^{N_u} S\left ( u_j,u_k \right )\\
=& e^{i p\left ( u_j \right )L} \prod_{\substack{k=1\\k\ne j}}^{N_u} S\left ( u_j,u_k \right )\\
=&: e^{i \phi_{u_j}}.
\end{align}
The Gaudin matrix is an $N_{u} \times N_{u}$ matrix whose elements are given by $G_{ij}=\partial_{u_i}\phi_{u_j}$, where
\begin{align}
\partial_{u_i}\phi_{u_j}&=\delta _{ij}\left ({p}'\left ( u_i \right ) L+
\sum_{k=1}^{N_u} \varphi\left ( u_i,u_k \right ) \right ) -\varphi\left ( u_i,u_j \right ),\\
{p}'\left ( u \right )&=\frac{-1}{u^2+\frac{1}{4}} , \\
\varphi\left ( u,v \right ) &=-i\frac{\partial }{\partial u} \log S\left ( u,v \right )=\frac{2}{1+(u-v)^{2}}  .
\end{align}
Notice that $\varphi(u,v)$ only depends on the absolute value of $(u-v)$.

The explicit expression of the factorized Gaudin determinant varies depending on whether the number of Bethe roots is even or odd.

\textbf{$N_u$ even.} We take the rapidities ordered as
\begin{align}
\{\mathbf{u}\}=\left \{ \mathbf{u}^{+} \right \}\cup \left \{ \mathbf{u}^{-}\right \}
=\left \{ u_1,u_2,\dots,u_{\frac{N_u}{2}} \right \}\cup \left \{ -u_1,-u_2,\dots,-u_{\frac{N_u}{2}} \right \}.\label{evenorder}
\end{align}
Gaudin determinant can be factorized as
\begin{align}
\det G=\det G_{+}^{even}\det G_{-}^{even},
\end{align}
where $G_{\pm}^{even}$ are $\frac{N_u}{2} \times \frac{N_u}{2}$ matrices whose elements are given by
\begin{align}
\left [ G_{\pm}^{even} \right ] _{ij}&=\delta _{ij}\left ( {p}'\left ( u_i \right ) L
+\sum_{k=1}^{\frac{N_u}{2}} \varphi^{+}\left ( u_i,u_k \right )  \right ) 
-\varphi ^{\pm}\left ( u_i,u_j \right ) ,\\
&\varphi ^{\pm}\left ( u_i,u_j \right )=\varphi\left ( u_i,u_j \right )\pm \varphi\left ( u_i,-u_j \right ). 
\end{align}

\textbf{$N_u$ odd.} Rapidities are ordered as
\begin{align}
\{\mathbf{u}\}=\left \{ \mathbf{u}^{+},0\right \}\cup \left \{ \mathbf{u}^{-}\right \}
=\left \{ u_1,u_2,\dots,u_{\frac{N_u}{2}},0 \right \}\cup \left \{ -u_1,-u_2,\dots,-u_{\frac{N_u}{2}} \right \}.\label{oddorder}
\end{align}
Gaudin determinant can be factorized as
\begin{align}
\det G=\det G_{+}^{odd}\det G_{-}^{odd},
\end{align}
where $G_{+}^{odd}$ is a $\left ( \left \lfloor \frac{N_u}{2} \right \rfloor +1 \right ) 
\times \left ( \left \lfloor \frac{N_u}{2} \right \rfloor +1 \right )$ matrix and $G_{-}^{odd}$ is a $\left \lfloor \frac{N_u}{2} \right \rfloor \times \left \lfloor \frac{N_u}{2} \right \rfloor$ matrix. $\left [ G_{\pm}^{odd} \right ]_{ij}, \ i,j\le \left \lfloor \frac{N_u}{2}\right \rfloor $ has a similar expression as $\left [ G_{\pm}^{even} \right ]_{ij}$
\begin{align}
\left [ G_{\pm}^{odd} \right ] _{ij}&=\delta _{ij}\left ( {p}'\left ( u_i \right ) L
+\sum_{k=1}^{\left \lfloor \frac{N_u}{2} \right \rfloor } \varphi^{+}\left ( u_i,u_k \right ) +\varphi\left ( u_i,0 \right )  \right ) 
-\varphi ^{\pm}\left ( u_i,u_j \right ) ,\\
&\varphi ^{\pm}\left ( u_i,u_j \right )=\varphi\left ( u_i,u_j \right )\pm \varphi\left ( u_i,-u_j \right ),
\end{align}

Other elements of $G_{+}^{odd}$ are
\begin{align}
\left [ G_{+}^{odd} \right ]_{\left \lfloor \frac{N_u}{2}\right \rfloor+1,j}
&= G_{\left \lfloor \frac{N_u}{2}\right \rfloor+1,j}, \ 
j  = 1,\dots,\left \lfloor \frac{N_u}{2}\right \rfloor+1,\\
\left [ G_{+}^{odd} \right ]_{i,\left \lfloor \frac{N_u}{2}\right \rfloor+1} 
&= 2 G_{i,\left \lfloor \frac{N_u}{2}\right \rfloor+1}, \ 
i  = 1,\dots,\left \lfloor \frac{N_u}{2}\right \rfloor.
\end{align}

The norm of an on-shell Bethe state in the $SU(2)$ sector with $N_u \ne 0$ takes the form \cite{gaudin1981normalization, Korepin:1982gg}

\begin{align}
\left \langle \mathbf{u}|\mathbf{u} \right \rangle
=\left ( \prod_{i<j} \frac{S\left ( u_i,u_j \right )}{S\left ( u_i^*,u_j^* \right )} \right )^{\frac{1}{2}} \prod_{k=1}^{N_u} \frac{1}{{p}'\left ( u_k \right)} \det G. \label{su2norm}
\end{align}
In the case of parity-invariant Bethe roots, the norm further simplifies depending on whether $N_u$ is even or odd:
\begin{align}
\left \langle \mathbf{u}|\mathbf{u} \right \rangle
=
\begin{cases}
\prod_{i=1}^{\frac{N_u}{2}}\left (\frac{S\left ( u_i,-u_i \right )}{S\left ( u_i^{*},-u_i^{*} \right )} \right )^{\frac{1}{2}}
\prod_{l>k=1}^{\frac{N_u}{2}} \left (\frac{S\left ( u_l,-u_k \right )}{S\left ( u_l^{*},-u_k^{*} \right )} \right )\\
\ \times \prod_{j=1}^{N_u/2}
\left ( \frac{1}{{p}'\left ( u_j \right)} \right)^2 
\det G_{+}^{even}\det G_{-}^{even} 
& N_u\ \text{even},  \\
\prod_{i=1}^{\left \lfloor \frac{N_u}{2} \right \rfloor }\left (\frac{S\left ( u_i,-u_i \right )}{S\left ( u_i^{*},-u_i^{*} \right )}\right )^{\frac{1}{2}}
\left (\frac{S\left ( u_i,0 \right )}{S\left ( u_i^{*},0 \right )}    \right )\prod_{l>k=1}^{\left \lfloor \frac{N_u}{2} \right \rfloor } \left (\frac{S\left ( u_l,-u_k \right )}{S\left ( u_l^{*},-u_k^{*} \right )}   \right ) \\
\ \times \frac{1}{{p}'\left (0\right)}\prod_{j=1}^{\left \lfloor N_u/2 \right \rfloor } 
\left ( \frac{1}{{p}'\left ( u_j \right)} \right)^2  
\det G_{+}^{odd}\det G_{-}^{odd}  
& N_u\ \text{odd}.
\end{cases}
\end{align}
Note that the Bethe roots are ordered as $\eqref{evenorder}, \eqref{oddorder}$.

In the $SU(2) \times SU(2)$ sector, the norm of on-shell Bethe state with $(N_u, N_w, N_v) = \left(\frac{L}{2}, 0, \frac{L}{2}\right)$ is given by product of $\eqref{su2norm}$
\begin{align}
\left \langle \mathbf{u},\mathbf{v}|\mathbf{u},\mathbf{v} \right \rangle
=&\left ( \prod_{i<j} \frac{S\left ( u_i,u_j \right )}{S\left ( u_i^*,u_j^* \right )} \right )^{\frac{1}{2}} 
\left ( \prod_{k<l} \frac{S\left ( v_k,v_l \right )}{S\left ( v_k^*,v_l^* \right )} \right )^{\frac{1}{2}} 
\prod_{m=1}^{N_u} \frac{1}{{p}'\left ( u_m \right)} 
\prod_{n=1}^{N_v} \frac{1}{{p}'\left ( v_n \right)} \nonumber\\
& \times \det G\left ( u \right )  \det G\left ( v \right ) .
\end{align}

\subsection{Overlap formulas}\label{subsection overlap}

The first exact results for the overlap formulas between the N\'eel state and Bethe eigenstates in the XXX spin chain were presented in \cite{pozsgay2014overlaps}, and were subsequently simplified into a factorized form containing Gaudin determinant and rigorously proved in \cite{brockmann2014gaudin, brockmann2014neel}. This proof relies on an off-shell overlap formula originally derived in \cite{tsuchiya1998determinant}. An alternative proof, based on the analytic properties in the coordinate Bethe ansatz formulation, was later provided in \cite{Jiang:2020sdw}. More recently, a general proof based on the algebraic Bethe ansatz was given in \cite{Gombor:2021uxz}, where the KT relation plays a significant role.

In the following, we present a detailed explanation of the overlap formulas for the first two chiral boundary states listed in \autoref{eight} and summarize the results for the remaining ones.

Boundary states of the form $\left \langle B \right |=\left \langle 2\bar{4}1\bar{4} \right |^{\otimes \frac{L}{2}}$ have non-vanishing overlaps only with Bethe states whose quantum numbers are $(N_u, N_w, N_v) = \left(\frac{L}{2}, 0, 0\right)$. The overlap is given by the wavefunction component associated with the state $\left|2\bar{4}1\bar{4}\right\rangle$ within the Bethe eigenstate:
\begin{align}
\left \langle B_{2\bar{4}1\bar{4},\frac{L}{2}} | \mathbf{u}\right \rangle 
=&\Psi_{2\bar{4}1\bar{4}}(\mathbf{u},L)\\
=&\sum_{\sigma \in S_{\frac{L}{2}}} \prod_{j=1}^{\frac{L}{2}} \left(l_{u_{\sigma_j}}\right)^{2j-1}
\prod_{\substack{j>k \\ \sigma_j<\sigma_k}}S\left ( u_{\sigma_j},u_{\sigma_k} \right )\label{2414wavefuction} \\
=&\left ( \prod_{j>k} f^{-1}\left ( u_j,u_k \right ) \right ) \sum_{\sigma \in S_{\frac{L}{2}}} \prod_{j=1}^{\frac{L}{2}} \left(l_{u_{\sigma_j}}\right)^{2j-1}
\prod_{j>k}f\left ( u_{\sigma_j},u_{\sigma_k} \right ).
\end{align}
When the Bethe state is on-shell and the rapidities are parity-invariant, this expression simplifies to a compact formula involving the Gaudin determinant \cite{brockmann2014neel}
\begin{align}
\label{b2414udetgplus}\left \langle B_{2\bar{4}1\bar{4},\frac{L}{2}} | \mathbf{u}\right \rangle=&
\prod_{j>k=1}^{N_u} f^{-1}\left ( u_j,u_k \right )\times\\ \nonumber
&\begin{cases}
\prod_{i=1}^{\frac{N_u}{2}}\frac{\left ( u_i^2+\frac{1}{4} \right )^2 }{4 u_i^2}
\prod_{j>k=1}^{\frac{N_u}{2}}F\left ( u_j,u_k \right )\times \det G_{+}^{even }  
  & N_u=\frac{L}{2}\ \text{even}, \\
\frac{1}{8}
\prod_{i=1}^{\left \lfloor \frac{N_u}{2} \right \rfloor }
\frac{\left ( u_i^2+\frac{1}{4} \right )^2 }{4 u_i^2}\frac{u_i^2+1}{u_i^2} 
\prod_{j>k=1}^{\left \lfloor \frac{N_u}{2} \right \rfloor}
F\left ( u_j,u_k \right )\times \det G_{+}^{odd}
  & N_u=\frac{L}{2}\ \text{odd}, 
\end{cases}
\end{align}
where
\begin{align}
    F\left ( u_j,u_k \right )=f\left ( u_j,u_k \right )f\left ( u_j,-u_k \right )
  f\left ( u_k,u_j \right )f\left ( -u_k,u_j \right ).
\end{align}
Thus, the overlap between the boundary state and the normalized Bethe state is given by
\begin{align}\label{nomalizedoverlapu}
\frac{\left \langle B_{2\bar{4}1\bar{4},\frac{L}{2}} | \mathbf{u}\right \rangle}
{\sqrt{\left \langle \mathbf{u} | \mathbf{u} \right \rangle }}=
&\begin{cases}
\prod_{i=1}^{\frac{N_u}{2}}\left ( \frac{S\left ( u_{i}^{*},-u_{i}^{*} \right ) }{S\left ( u_{i},-u_{i} \right )}  \right )^{\frac{1}{4}}
\prod_{j>k=1}^{\frac{N_u}{2}}\frac{S\left ( u_{j},-u_{k} \right )}{\left | S\left ( u_{j},-u_{k} \right ) \right | } \\
\ \ \ \times \prod_{i=1}^{\frac{N_u}{2}}\frac{ u_i-\frac{i}{2} }{4 u_i}
\sqrt{\frac{\det G_{+}^{even }}{\det G_{-}^{even }} }   
  & N_u=\frac{L}{2}\ \text{even}, \\
\frac{1}{4}
\prod_{i=1}^{\left \lfloor \frac{N_u}{2} \right \rfloor}
\left ( \frac{S\left ( u_{i}^{*},-u_{i}^{*} \right ) }{S\left ( u_{i},-u_{i} \right )}  \right )^{\frac{1}{4}}
\frac{S\left ( u_{i},0 \right )}{\left | S\left ( u_{i},0 \right ) \right | }
\prod_{j>k=1}^{\left \lfloor \frac{N_u}{2} \right \rfloor}\frac{S\left ( u_{j},-u_{k} \right )}{\left | S\left ( u_{j},-u_{k} \right ) \right | } \\
\ \ \ \times \prod_{i=1}^{\left \lfloor \frac{N_u}{2} \right \rfloor }
\frac{u_i-\frac{i}{2} }{4 u_i}
\sqrt{\frac{\det G_{+}^{odd}}{\det G_{-}^{odd}} } 
  & N_u=\frac{L}{2}\ \text{odd}.
\end{cases}
\end{align}
The Bethe roots are arranged according to the ordering in $\eqref{evenorder}, \eqref{oddorder}$.

For $\left \langle B \right |=\left \langle 1\bar{4}2\bar{4} \right | ^{\otimes \frac{L}{2} }$, the results are basically the same as $\left \langle B \right |=\left \langle 2\bar{4}1\bar{4} \right |^{\otimes \frac{L}{2}}$. 
\begin{align}
\left \langle B_{1\bar{4}2\bar{4},\frac{L}{2}} | \mathbf{u}\right \rangle
=&\sum_{\sigma \in S_{\frac{L}{2}}} \prod_{j=1}^{\frac{L}{2}} \left(l_{u_{\sigma_j}}\right)^{2j}
\prod_{\substack{j>k \\ \sigma_j<\sigma_k}}S\left ( u_{\sigma_j},u_{\sigma_k} \right ), \\
=&\left ( \prod_{j>k} f^{-1}\left ( u_j,u_k \right ) \right ) \sum_{\sigma \in S_{\frac{L}{2}}} \prod_{j=1}^{\frac{L}{2}} \left(l_{u_{\sigma_j}}\right)^{2j}
\prod_{j>k}f\left ( u_{\sigma_j},u_{\sigma_k} \right ).
\end{align}
By comparison with $\eqref{2414wavefuction}$, we deduce that the final result is identical to $\eqref{b2414udetgplus}$ and $\eqref{nomalizedoverlapu}$, up to an overall multiplicative factor
\begin{align}
\prod_{j = 1}^{N_u} l_{u_j} = (-1)^{N_u}=(-1)^{\frac{L}{2}}. %\begin{cases}
%1  & N_u = \frac{L}{2}\ \text{even}, \\
%-1  & N_u = \frac{L}{2}\ \text{odd}.
%\end{cases}
\end{align}
Here the pair condition on the Bethe roots $\mathbf{u}$ has been used. 
The remaining results are as follows. For the quantum numbers $(N_u, N_w, N_v) = \left(0, 0, \frac{L}{2}\right)$, the expression corresponding to $\left \langle B \right |=\left \langle 1\bar{3}1\bar{4} \right | ^{\otimes \frac{L}{2} }$ can be obtained from that of $\left \langle B \right |=\left \langle 2\bar{4}1\bar{4} \right | ^{\otimes \frac{L}{2} }$ by replacing $\mathbf{u}$ with $\mathbf{v}$. Likewise, the result for $\left \langle B \right |=\left \langle 1\bar{4}1\bar{3} \right | ^{\otimes \frac{L}{2} }$ is derived from that of $\left \langle B \right |=\left \langle 1\bar{4}2\bar{4} \right | ^{\otimes \frac{L}{2} }$ under the same substitution $\mathbf{u} \rightarrow \mathbf{v}$. In the $SU(2) \times SU(2)$ sector characterized by the quantum numbers $(N_u, N_w, N_v) = \left(\frac{L}{2}, 0, \frac{L}{2}\right)$, the corresponding results are given below:

\begin{align}
\left \langle B_{2\bar{3}1\bar{4},\frac{L}{2}} | \mathbf{u}, \mathbf{v}\right \rangle
&=\left \langle B_{2\bar{4}1\bar{4},\frac{L}{2}} | \mathbf{u}\right \rangle
\times \left \langle B_{1\bar{3}1\bar{4},\frac{L}{2}} | \mathbf{v}\right \rangle ,\\
\left \langle B_{2\bar{4}1\bar{3},\frac{L}{2}} | \mathbf{u}, \mathbf{v}\right \rangle
&=\left \langle B_{2\bar{4}1\bar{4},\frac{L}{2}} | \mathbf{u}\right \rangle
\times \left \langle B_{1\bar{4}1\bar{3},\frac{L}{2}} | \mathbf{v}\right \rangle ,\\
\left \langle B_{1\bar{3}2\bar{4},\frac{L}{2}} | \mathbf{u}, \mathbf{v}\right \rangle
&=\left \langle B_{1\bar{4}2\bar{4},\frac{L}{2}} | \mathbf{u}\right \rangle
\times \left \langle B_{1\bar{3}1\bar{4},\frac{L}{2}} | \mathbf{v}\right \rangle ,\\
\left \langle B_{1\bar{4}2\bar{3},\frac{L}{2}} | \mathbf{u}, \mathbf{v}\right \rangle
&=\left \langle B_{1\bar{4}2\bar{4},\frac{L}{2}} | \mathbf{u}\right \rangle
\times \left \langle B_{1\bar{4}1\bar{3},\frac{L}{2}} | \mathbf{v}\right \rangle.
\end{align}

\section{Conclusion} \label{section conclusion}

In this paper, we studied chiral integrable boundary states in the ABJM spin chain. These states had not appeared in previous investigations of integrable boundary states within the $AdS_4/CFT_3$ correspondence. Among the basis states, we identified  specific two-site or four-site states satisfy the untwisted integrable condition and exhibit this chiral property. We have some numerical evidences that other basis states are not chiral integrable.  We further computed the overlap between these chiral integrable states and general Bethe states.

A key step in our analysis was the use of~\eqref{omn=obmn} which is a sufficient condition for a state to be chiral integrable. Crucially, this sufficient condition is simpler to solve in practice than~\eqref{ohs=ohbs}—which is equivalent to the untwisted integrable condition—or than the formulation involving local higher charges generated from the transfer matrices $\tau(u)$ and $\bar\tau(u)$. Since numerical results support that other basis states are not chiral integrable, we conjecture that this sufficient condition is also necessary. If one can proof this, then the determination of all chiral integrable states among a certain set of states will be greatly simplified.   We also anticipate discovering more complex chiral integrable states, such as matrix product states (MPS), based on solutions to the corresponding KT relation~\cite{Gombor:2020kgu}. An alternative approach involves generalizing the necessary integrability condition for MPS~\cite{deLeeuw:2024qki} from $SO(N)$ spin chains to the ABJM setting. This condition allows us to systematically identify candidate states. Then we can try to determine genuine integrable states through analytical or numerical methods.

It would also be valuable to explore how these chiral integrable states might be realized as operators or defects in ABJM theory and to seek their interpretation in the dual string/M-theory. Notably, known integrable boundary states in $AdS_4/CFT_3$ correspondence arising from BPS operators or defects are exclusively achiral. This presents a stark contrast to ${\cal N}=4$ SYM theory, where both chiral and achiral integrable states in the $SO(6)$ sector emerge in computations of specific correlators. For example, chiral states appear in heavy-heavy-light (HHL) three-point functions~\cite{Jiang:2019xdz, Jiang:2019zig}, while achiral states feature in domain-wall one-point functions~\cite{deLeeuw:2015hxa}. Understanding the deeper underlying reasons for this difference between the two theories merits further investigation.

\section*{Acknowledgments}

It is our pleasure to thank Yunfeng Jiang and Yu-Xuan Zhang for very helpful discussions. This work is supported  by the National Natural Science Foundation of China (NSFC) Grants No.~12375006, 11975164,  12247103, 11935009, and 
Tianjin University Self-Innovation Fund Extreme Basic Research Project Grant No.~2025XJ21-0007.
Some numerical calculations were conducted on the CJQS-HPC platform at Tianjin University.

\appendix
\section{Bethe roots}\label{bethe root}
Solving rational Q-system  \cite{Marboe:2016yyn, Gu:2022dac} of given quantum numbers $(L,N_u,N_w,N_v)$ yields $\left \{ Q_1(u), Q_2(w), Q_3(v) \right \}$ whose zeros are Bethe roots.

\begin{table}[H]
\centering
\begin{tabular}{|c|c|c|c|c|}
\hline
\multirow{5}{*}{\makecell{$SU(2)$ \\ $\left(N_u,N_w,N_v\right )=\left ( \frac{L}{2}, 0, 0\right )$}}     & L                   & 2                  & \multicolumn{1}{c|}{4}                   & 6   \\ \cline{2-5}
& \multirow{4}{*}{$Q_1(u)$} & \multirow{4}{*}{$u$} & \multicolumn{1}{c|}{\multirow{4}{*}{$u^2-\frac{1}{12} $}} & $u^3+\frac{u}{12}-\frac{1}{4 \sqrt{3}}$ \\ \cline{5-5}
&                     &                    & \multicolumn{1}{c|}{}                    & $u^3+\frac{u}{12}+\frac{1}{4 \sqrt{3}}$ \\ \cline{5-5} 
&                     &                    & \multicolumn{1}{c|}{}                    & $u^3+\frac{1}{12} \left(5-2 \sqrt{13}\right) u$ \\ \cline{5-5} 
 &                     &                    & \multicolumn{1}{c|}{}                    & $u^3+\frac{1}{12} \left(2 \sqrt{13}+5\right) u$ \\ \hline
\multirow{2}{*}{\makecell{$SU(2) \times SU(2)$ \\ $\left (N_u,N_w,N_v\right )=\left (\frac{L}{2},0,\frac{L}{2} \right )$}} & L                   & 2                  & \multicolumn{2}{c|}{4}                         \\ \cline{2-5} 
& $\left \{ Q_1(u), Q_3(v) \right \}$             & $\left \{ u,v \right \}$           & \multicolumn{2}{c|}{$\left \{ u^2-\frac{1}{12},v^2-\frac{1}{12} \right \}$}                 \\ \hline
\end{tabular}
\caption{Bethe roots}
\end{table}

Several Bethe roots don`t exhibit chiral pair structure are also used in our numerical check, which are: 
\begin{itemize}
\item $L=4n-2\, (n=1, 2, \cdots), \,N_u=N_w=N_v=1,\, {\bf u}={\bf v}={\bf w}=\{\pm1/2\}$,

\item $L=3n-1,( n=1, 2, \cdots), \, N_{ u}=N_{w}=N_{v}=1, \, {\bf u}=-{\bf v}=\{1/(2\sqrt{3})\},\, {\bf w}=\{0\}$,

\item $L=3, \, N_{u}=N_{w}=1, \,N_{v}=2,\, {\bf u}={\bf w}=\{\pm\frac{\sqrt{3}}{2}\}, {\bf v}=\{\frac{1}{8}(\pm \sqrt{3}+\sqrt{11}), \frac{1}{8}(\pm \sqrt{3}-\sqrt{11}) \}$,~\footnote{This set of Bethe roots was first found numerically in \cite{Jiang:2023cdm}. We now find the analytic expression.}

\item Bethe roots listed in the appendixes of \cite{Yang:2021hrl, Jiang:2023cdm} besides the above ones.

\end{itemize}


\begin{thebibliography}{10}

\bibitem{Ghoshal:1993tm}
Subir Ghoshal and Alexander~B. Zamolodchikov.
\newblock {Boundary S matrix and boundary state in two-dimensional integrable quantum field theory}.
\newblock {\em Int. J. Mod. Phys. A}, 9:3841--3886, 1994.
\newblock [Erratum: Int.J.Mod.Phys.A 9, 4353 (1994)].

\bibitem{Piroli:2017sei}
Lorenzo Piroli, Bal\'azs Pozsgay, and Eric Vernier.
\newblock {What is an integrable quench?}
\newblock {\em Nucl. Phys. B}, 925:362--402, 2017.

\bibitem{Gombor:2020kgu}
Tamas Gombor and Zoltan Bajnok.
\newblock {Boundary states, overlaps, nesting and bootstrapping AdS/dCFT}.
\newblock {\em JHEP}, 10:123, 2020.

\bibitem{deLeeuw:2015hxa}
Marius de~Leeuw, Charlotte Kristjansen, and Konstantin Zarembo.
\newblock {One-point Functions in Defect CFT and Integrability}.
\newblock {\em JHEP}, 08:098, 2015.

\bibitem{Buhl-Mortensen:2015gfd}
Isak Buhl-Mortensen, Marius de~Leeuw, Charlotte Kristjansen, and Konstantin Zarembo.
\newblock {One-point Functions in AdS/dCFT from Matrix Product States}.
\newblock {\em JHEP}, 02:052, 2016.

\bibitem{deLeeuw:2016umh}
Marius de~Leeuw, Charlotte Kristjansen, and Stefano Mori.
\newblock {AdS/dCFT one-point functions of the SU(3) sector}.
\newblock {\em Phys. Lett. B}, 763:197--202, 2016.

\bibitem{deLeeuw:2017dkd}
Marius de~Leeuw, Asger~C. Ipsen, Charlotte Kristjansen, Kasper~E. Vardinghus, and Matthias Wilhelm.
\newblock {Two-point functions in AdS/dCFT and the boundary conformal bootstrap equations}.
\newblock {\em JHEP}, 08:020, 2017.

\bibitem{Buhl-Mortensen:2017ind}
Isak Buhl-Mortensen, Marius de~Leeuw, Asger~C. Ipsen, Charlotte Kristjansen, and Matthias Wilhelm.
\newblock {Asymptotic One-Point Functions in Gauge-String Duality with Defects}.
\newblock {\em Phys. Rev. Lett.}, 119(26):261604, 2017.

\bibitem{DeLeeuw:2018cal}
Marius De~Leeuw, Charlotte Kristjansen, and Georgios Linardopoulos.
\newblock {Scalar one-point functions and matrix product states of AdS/dCFT}.
\newblock {\em Phys. Lett. B}, 781:238--243, 2018.

\bibitem{DeLeeuw:2019ohp}
Marius De~Leeuw, Tam\'as Gombor, Charlotte Kristjansen, Georgios Linardopoulos, and Bal\'azs Pozsgay.
\newblock {Spin Chain Overlaps and the Twisted Yangian}.
\newblock {\em JHEP}, 01:176, 2020.

\bibitem{Kristjansen:2020mhn}
Charlotte Kristjansen, Dennis M\"uller, and Konstantin Zarembo.
\newblock {Integrable boundary states in D3-D5 dCFT: beyond scalars}.
\newblock {\em JHEP}, 08:103, 2020.

\bibitem{Jiang:2019xdz}
Yunfeng Jiang, Shota Komatsu, and Edoardo Vescovi.
\newblock {Structure constants in $ \mathcal{N} $ = 4 SYM at finite coupling as worldsheet g-function}.
\newblock {\em JHEP}, 07(07):037, 2020.

\bibitem{Jiang:2019zig}
Yunfeng Jiang, Shota Komatsu, and Edoardo Vescovi.
\newblock {Exact Three-Point Functions of Determinant Operators in Planar $N=4$ Supersymmetric Yang-Mills Theory}.
\newblock {\em Phys. Rev. Lett.}, 123(19):191601, 2019.

\bibitem{Kristjansen:2023ysz}
Charlotte Kristjansen and Konstantin Zarembo.
\newblock {\textquoteright{}t Hooft loops and integrability}.
\newblock {\em JHEP}, 08:184, 2023.

\bibitem{Gombor:2024api}
Tamas Gombor and Zolt\'an Bajnok.
\newblock {Dual overlaps and finite coupling \textquoteright{}t Hooft loops}.
\newblock {\em JHEP}, 12:034, 2024.

\bibitem{Kristjansen:2024map}
Charlotte Kristjansen and Konstantin Zarembo.
\newblock {\textquoteright{}t Hooft loops in N=4 super-Yang-Mills}.
\newblock {\em JHEP}, 02:179, 2025.

\bibitem{Holguin:2025bfe}
Adolfo Holguin and Hiroki Kawai.
\newblock {Integrability and Conformal Blocks for Surface Defects in $\mathcal{N}=4$ SYM}.
\newblock 3 2025.

\bibitem{Chalabi:2025nbg}
Adam Chalabi, Charlotte Kristjansen, and Chenliang Su.
\newblock {Integrable corners in the space of Gukov-Witten surface defects}.
\newblock {\em Phys. Lett. B}, 866:139512, 2025.

\bibitem{Coronado:2025xwk}
Frank Coronado, Shota Komatsu, and Konstantin Zarembo.
\newblock {Coulomb Branch and Integrability}.
\newblock 6 2025.

\bibitem{Yang:2021hrl}
Peihe Yang, Yunfeng Jiang, Shota Komatsu, and Jun-Bao Wu.
\newblock {Three-point functions in ABJM and Bethe Ansatz}.
\newblock {\em JHEP}, 01:002, 2022.

\bibitem{Kristjansen:2021abc}
Charlotte Kristjansen, Dinh-Long Vu, and Konstantin Zarembo.
\newblock {Integrable domain walls in ABJM theory}.
\newblock {\em JHEP}, 02:070, 2022.

\bibitem{Gombor:2022aqj}
Tamas Gombor and Charlotte Kristjansen.
\newblock {Overlaps for matrix product states of arbitrary bond dimension in ABJM theory}.
\newblock {\em Phys. Lett. B}, 834:137428, 2022.

\bibitem{Yang:2022dlk}
Peihe Yang.
\newblock {Integrable boundary states from maximal giant gravitons in ABJM theory}.
\newblock {\em Phys. Lett. B}, 846:138194, 2023.

\bibitem{Jiang:2023cdm}
Yunfeng Jiang, Jun-Bao Wu, and Peihe Yang.
\newblock {Wilson-loop one-point functions in ABJM theory}.
\newblock {\em JHEP}, 09:047, 2023.

\bibitem{Wu:2024uix}
Jun-Bao Wu and Peihe Yang.
\newblock {Three-point functions in Aharony-Bergman-Jafferis-Maldacena theory and integrable boundary states}.
\newblock {\em JHEP}, 02:030, 2025.

\bibitem{Bai:2024qtg}
Nan Bai and Mao-Zhong Shao.
\newblock {Integrable matrix product states of ABJM theory from projecting method}.
\newblock {\em Mod. Phys. Lett. A}, 40(19n20):2550068, 2025.

\bibitem{Linardopoulos:2022wol}
Georgios Linardopoulos.
\newblock {String integrability of the ABJM defect}.
\newblock {\em JHEP}, 06:033, 2022.

\bibitem{Linardopoulos:2021rfq}
Georgios Linardopoulos and Konstantin Zarembo.
\newblock {String integrability of defect CFT and dynamical reflection matrices}.
\newblock {\em JHEP}, 05:203, 2021.

\bibitem{Linardopoulos:2025ypq}
Georgios Linardopoulos.
\newblock {String theory methods for defect CFTs}.
\newblock 1 2025.

\bibitem{Aharony:2008ug}
Ofer Aharony, Oren Bergman, Daniel~Louis Jafferis, and Juan Maldacena.
\newblock {N=6 superconformal Chern-Simons-matter theories, M2-branes and their gravity duals}.
\newblock {\em JHEP}, 10:091, 2008.

\bibitem{Minahan:2008hf}
J.~A. Minahan and K.~Zarembo.
\newblock {The Bethe ansatz for superconformal Chern-Simons}.
\newblock {\em JHEP}, 09:040, 2008.

\bibitem{Bak:2008cp}
Dongsu Bak and Soo-Jong Rey.
\newblock {Integrable Spin Chain in Superconformal Chern-Simons Theory}.
\newblock {\em JHEP}, 10:053, 2008.

\bibitem{Marboe:2016yyn}
Christian Marboe and Dmytro Volin.
\newblock {Fast analytic solver of rational Bethe equations}.
\newblock {\em J. Phys. A}, 50(20):204002, 2017.

\bibitem{Gu:2022dac}
Jie Gu, Yunfeng Jiang, and Marcus Sperling.
\newblock {Rational $Q$-systems, Higgsing and mirror symmetry}.
\newblock {\em SciPost Phys.}, 14(3):034, 2023.

\bibitem{Pronko:1998xa}
G.~P. Pronko and Yu.~G. Stroganov.
\newblock {Bethe equations 'on the wrong side of equator'}.
\newblock {\em J. Phys. A}, 32:2333--2340, 1999.

\bibitem{Gromov:2007ky}
Nikolay Gromov and Pedro Vieira.
\newblock {Complete 1-loop test of AdS/CFT}.
\newblock {\em JHEP}, 04:046, 2008.

\bibitem{brockmann2014neel}
Michael Brockmann, Jacopo De~Nardis, Bram Wouters, and Jean-S{\'e}bastien Caux.
\newblock N{\'e}el-xxz state overlaps: odd particle numbers and lieb--liniger scaling limit.
\newblock {\em Journal of Physics A: Mathematical and Theoretical}, 47(34):345003, 2014.

\bibitem{Jiang:2020sdw}
Yunfeng Jiang and Bal\'azs Pozsgay.
\newblock {On exact overlaps in integrable spin chains}.
\newblock {\em JHEP}, 06:022, 2020.

\bibitem{gaudin1981normalization}
Michel Gaudin, Barry~M McCoy, and Tai~Tsun Wu.
\newblock Normalization sum for the bethe's hypothesis wave functions of the heisenberg-ising chain.
\newblock {\em Physical Review D}, 23(2):417, 1981.

\bibitem{Korepin:1982gg}
V.~E. Korepin.
\newblock {CALCULATION OF NORMS OF BETHE WAVE FUNCTIONS}.
\newblock {\em Commun. Math. Phys.}, 86:391--418, 1982.

\bibitem{pozsgay2014overlaps}
Bal{\'a}zs Pozsgay.
\newblock Overlaps between eigenstates of the xxz spin-1/2 chain and a class of simple product states.
\newblock {\em Journal of Statistical Mechanics: Theory and Experiment}, 2014(6):P06011, 2014.

\bibitem{brockmann2014gaudin}
Michael Brockmann, Jacopo De~Nardis, Bram Wouters, and Jean-S{\'e}bastien Caux.
\newblock A gaudin-like determinant for overlaps of n{\'e}el and xxz bethe states.
\newblock {\em Journal of Physics A: Mathematical and Theoretical}, 47(14):145003, 2014.

\bibitem{tsuchiya1998determinant}
Osamu Tsuchiya.
\newblock Determinant formula for the six-vertex model with reflecting end.
\newblock {\em arXiv preprint solv-int/9804010}, 1998.

\bibitem{Gombor:2021uxz}
Tam\'as Gombor and Bal\'azs Pozsgay.
\newblock {On factorized overlaps: Algebraic Bethe Ansatz, twists, and Separation of Variables}.
\newblock {\em Nucl. Phys. B}, 967:115390, 2021.

\bibitem{deLeeuw:2024qki}
Marius de~Leeuw and Adolfo Holguin.
\newblock {Integrable Conformal Defects in N=4 SYM}.
\newblock 6 2024.

\end{thebibliography}
\end{document}